\documentclass[onecolumn]{emulateapj}

\shorttitle{Faber-Jackson relation of galaxy clusters}
\shortauthors{C. E. Navia}

\usepackage{amsmath}

\begin{document}


\title
{
On the generalized Faber-Jackson relation for galaxy clusters

}

\author{Carlos. E. Navia }
\affil{Instituto de F\'{i}sica, Universidade Federal Fluminense, 24210-346, Niter\'{o}i, RJ, Brazil}


\altaffiltext{1}{E-mail address:navia@if.uff.br}


\begin{abstract}
The significant deviations among observations and the expectations based on self-similar scaling model of galaxy clusters, especially up to redshift $z\lesssim 0.4$, constrain the evolution of the X-ray clusters scaling relations with the redshift, is claimed that in this redshift range, the data has a strong influence by selection bias. However,
also suggests that some non-gravitational processes can be responsible for a weak or almost null evolution, at least to $z\lesssim 0.4$.
This almost universality observed in X-ray galaxy clusters can be understood if we assume that the X-ray emission, results from thermal bremsstrahlung from a hot diffuse intracluster gas
with temperatures about $10^8$ K. A fraction of it would not be bound to the cluster and would escape as a wind. This hot wind can warm the local environment, the thermal bath where the cluster is immersed. This mechanism can put all the galaxy clusters within thermal baths, with almost the same effective temperature, independent of the cluster redshift and it can be effective for clusters with redshifts up to $z\sim 0.4$. 
 Debye Gravitational Theory (DGT), allows obtaining a Generalized Faber-Jackson relation to described the galaxy clusters such as the M-$\sigma$ and M-Tx relations as a function of the bath thermal temperature.
 We show that the DGT prediction to the M-$\sigma$ relation, overlap the fit on data of an extensive spectroscopic survey
of galaxy clusters with MMT/Hectospec, at 0.1$<z<$0.3. And the DGT predictions to the M-Tx relation almost overlap the fit on data from Canada–France–Hawaii Telescope Lensing Survey and XMM-CFHTLS surveys up to $z\sim 0.47$.
\end{abstract}



\keywords{galaxy clusters, extragalactic astrophysics, dark matter}


\section{Introduction}
\label{sec:intro}

Galaxy clusters are the largest gravitationally bound structures in the universe. They contain from some to thousands of galaxies of all ages, shapes, and sizes, together totaling a mass about $10^{14}-10^{15}$ times the Sun mass. A hot gas emitting X-rays fills the Galaxy clusters, the mass of the gas may exceed the mass of stars in cluster galaxies. 
 
From the observation of the motions of galaxies near the edge of the Coma cluster (also known as Abell 1656) in the 30's, Zwicky found that the mass cluster derived from the virial method greatly exceed that from visual inspection. There was not enough visible mass to explain these movements.  That is the source of the first ideas about the dark matter in the universe.

Besides, according to the Big Bang Cosmology only $\sim 4\%, $ of the Universe is constituted by ordinary matter, the so-called baryonic matter, forming the stars, gas, dust, and planets that we see. However, the cosmology required more $\sim 23\%$ of matter to take into account the observations, this percentage the matter which we do not see is known as dark matter.  
Even so, the dark matter still could be baryonic matter in the form of
frozen brown dwarfs or small, dense chunks of heavy elements. These possibilities are known as massive compact halo objects, or ``MACHOs'' \citep{grie93}.
However, the hypothesis the baryonic dark matter destroys an of the pillars of cosmology, 
the Big Bang Nucleosynthesis (BBN). The robustness of the BBN indicates
that if there is dark matter, it should not be from a baryonic origin.

Then if  dark matter is not baryonic at all,  it should be made up of other, more exotic 
particles like a heavy lepton or  WIMPS (Weakly Interacting Massive Particles).
So, the hunt for MACHOs has reduced, and start the hunt for WIMPs.
However, despite an exhaustive hunt for WIMPs, the results were so far negative \citep{apri18,tan16,aker16}, including
the results of a new experiment, COSINE-100 \cite{cosi10} that constrain the DAMA/LIBRA experiment results, that for 20 years have claimed to have direct evidence for dark matter.
So after years of failed search, the dark matter hunters are now focusing on a theoretical particle,  much lighter than the WIMP, the axion; thus the Axion Dark Matter Experiment began to run.
Also, the search of dark particles as super-symmetric particles at LHC (CERN) so far is also negative\footnote{\url{https://atlas.web.cern.ch/Atlas/GROUPS/PHYSICS/CombinedSummaryPlots/SUSY/}} .

The amount of dark matter in the Universe, constitute an of the six free parameters of the
$\Lambda$CDM (Lambda cold dark matter) model \citep{akra18}, it provides a good description of the Planck CMB observations. Even so, some tensions remain, such as the Hubble controversy.
After GAIA DR2, the tension in the Hubble constant between the local measurement 
(Gaia DR2 parallaxes and HST photometry) \citep{ries18}, and the Planck CMB measurement increases to 3.8 sigmas. 
The new data raises the current tension between the late and early Universe route to the Hubble constant.
 If this divergence is real, it means there's basic new physics going on.

The hypothesis of the Dark matter increases from the 70's, to account for the rotation of nearby spiral galaxies, which didn't seem to have enough baryonic matter to have flat rotation curves.
So far, most of the models to describe the galaxy clusters are from the concordance cosmological model or $\Lambda$ CDM and requires two possible elements, whose nature is not yet known, the dark energy and the dark matter. 

The main alternative to the dark matter is the so-called modified gravity theories, especially the non-relativistic theories such as the 
Modified Newtonian Dynamics (MOND) an empirically motivated modification of Newtonian gravity or inertia suggested by Milgrom in 1983 \citep{milg83a,milg83b}, the Moffat's theory (MOG) \citep{moff13} and the emergent gravity theory \citep{verl17}, among others. These non-relativistic theories explain the galaxies rotation curves in the nearby Universe (z$\sim 0$) without dark matter.
Especially MOND is very well successful to describe the galaxies dynamics \citep{mcga11,fama12,sand90,krou12}. Indeed, MOND predicted the Tully-Fisher relation, a scaling law between the mass and the circular velocity at large radii, observed in nearby spiral galaxies. 

However, from all modified theories only MOND has results for the several galaxy cluster relations, such as the mass-temperature relation, (according to \cite{dema17}, it is not clear if MOG can describe the galaxy cluster). Even so, is well known that the MOND has some limitations. The galaxy cluster seems to indicate that still is necessary a residual mass even in MOND, an exotic neutrino, the ``sterile neutrino'', was considered as a  promising candidate to the hot dark matter required by MOND \citep{sand07,angu08,angu11}. 
So far,  there is no direct evidence of these neutrinos  \citep{aart16,adam16}.  
Indeed, in the central part of clusters, the observed acceleration is usually slightly larger than $a_0$ \citep{bell03}.
This requirement suggests an increase of the acceleration scale in MOND, such as made in 
Extended MOND (EMOND) \citep{zhao12}.

In this paper, we present another alternative to describe the dynamic of galaxy clusters from DGT, a thermodynamic gravitational theory. It is an extension for low temperatures of the Entropic Gravity Theory (EGT) (Verlinde, 2011),
DGT is very well successfully describing the galaxies rotation curves in a wide range of redshift \citep{navi17,navi18a}, and the kinematic the dwarf galaxies \citep{navi18b}. Prediction from DGT for the rotation curves of galaxies at
high redshift is in agreement from those obtained from the falling rotation curves, observed by VLT telescope \citep{lang17}.

 The organization of this paper is as follow.
In section~\ref{DGT} we present a description of galaxy clusters within the Debye gravitational theory, including the generalized Faber-Jackson relation \ref{faber},  and their applications to obtain the $M-\sigma$ relation \ref{m_sigma}, the X-ray emission \ref{xray_emission}, and the M-Tx relation \ref{m_t} as a function of the temperature of the thermal bath where the galaxy clusters are immersed. 
The survey includes comparisons among DGT predictions and data from the literature, as well as, with other models. In section~\ref{toy} we present a toy model to obtain the caustic pattern in the galaxy cluster, and in section~\ref{conclusions} we present some discussions and conclusions.

\section{Galaxy clusters within Debye Gravitational Theory (DGT)}
\label{DGT}

vvvvIn 1912 Debye postulated that the behavior of the  specific heat of 
a solid a low temperatures is a consequence of the vibrations of the atoms of the lattice of the solid, following the analogy to the photons modes in a cavity (blackbody radiation). In the Debye theory there is a continuous range of frequencies that cuts off at a maximum frequency $\omega_D$, or temperature $T_D=\hbar \omega_D/k_B$, where $\hbar$ is the Planck constant and $k_B$ is the Boltzmann constant. The Debye temperature $T_D$ is characteristic of a particular solid. At high temperatures $T>>T_D$ the Debye theory coincides with the Law of Dulong and Petit, where the specific heat is constant, it does not depend on the temperature.

In the DGT picture, the Newton theory of gravity plays the role of Dulong-Petit law. Also, the Newton theory can be obtained from thermodynamic concepts, more specifically from the Entropic Gravity Theory (EGT) \citep{verl11}. In this sense, in DGT gravity is induced by the entropy variation of a system constituted by oscillating quasi-particles (information bits) on a closed holographic screen and that stores the information of matter enclosed within it \citep{navi17}. In short, DGT introduced the Debye scheme in the entropic gravity theory to explain gravity at low temperatures.

Following this scheme and under a spherical symmetry the main equations of DGT \citep{navi17,navi18b} are

 \begin{equation}
 a\mathcal{D}_1\left(\frac{T_D}{T}\right)=\frac{GM}{R^2},
 \label{eq_1}
 \end{equation}
where $\mathcal{D}_1$ is the Debye first function defined as
\begin{equation}
\mathcal{D}_1\left(\frac{T_D}{T}\right)=\frac{T}{T_0}\int_0^{T_0/T} \frac{x}{\exp{x}-1}dx,
\end{equation}

In the limit for $T_D \gg T$, and considering that the temperature is proportional to the acceleration (Unruh effect), we have the constrain condition $\mathcal{D}_1(T_D/T)=(\pi^2/6)\; T/T_D=a/a_0$, and constitute a bound between the Debye temperature $T_D$ and the acceleration scale $a_0$. So Eq.~\ref{eq_1} becomes
\begin{equation}
a\left(\frac{a}{a_0}\right)=\frac{GM}{R^2}.
\label{eq_3}
\end{equation}

This equation is the root of the MOND theory,known as the deep-MOND regime. In general (for all range the temperatures) the 
Eq.~\ref{eq_3} can be parametrized by a power function as
\begin{equation}
a\left(\frac{a}{a_0}\right)^{\alpha}= \frac{GM}{R^2}.
\label{eq:mainDGT}
\end{equation}
The two asymptotically  cases are: 

\[ \alpha =
  \begin{cases}
    0      & \quad \text{then } a=GM/R^2        \text{ (Newton-high T)}\\
    1      & \quad \text{then } a(a/a_0)=GM/R^2 \text{ (deep-MOND-low T)}.\\
  \end{cases}
\]

 All values possibles for $\alpha$ can be obtained from
 
  \begin{equation}
 \alpha = \frac{\log \mathcal{D}_1\left(\frac{a_0}{a}\right)}{\log \frac{a}{a_0}}.
 \label{alpha_a}
 \end{equation}
 Taking into account the bond between the acceleration and temperature $a/a_0=(\pi^2/6)T/T_D$, the index $\alpha$, can be written as a function of temperature.

 On the other hand, the CMB data, at least up to redshift of z$\sim$ 3, is consistent with a linear relation,between the temperature and redshift 
\begin{equation}
\frac{T}{T_0}=(1+z),
\end{equation}
where $T_0$ is the temperature at $z=0$, that is, the current temperature, $T_0=2,73\;K$. The last equation can be expressed as
\begin{equation}
\frac{T}{T_D}=\frac{T_0}{T_D} (1+z)=0.43 (1+z),
\end{equation}
where $T_D$ id the Debye temperature, $T_D=T_0/0.43=6.35\;K$. Thus, the only one free parameter of the DGT, the Debye temperaute in DGT is a little more than twice as much of  $T_0$. This last equation is an useful expression because allow obtaining $\alpha$ as a function of redshift.
\begin{figure}
\vspace*{-0.0cm}
\hspace*{0.0cm}
\centering
\includegraphics[width=10.0cm]{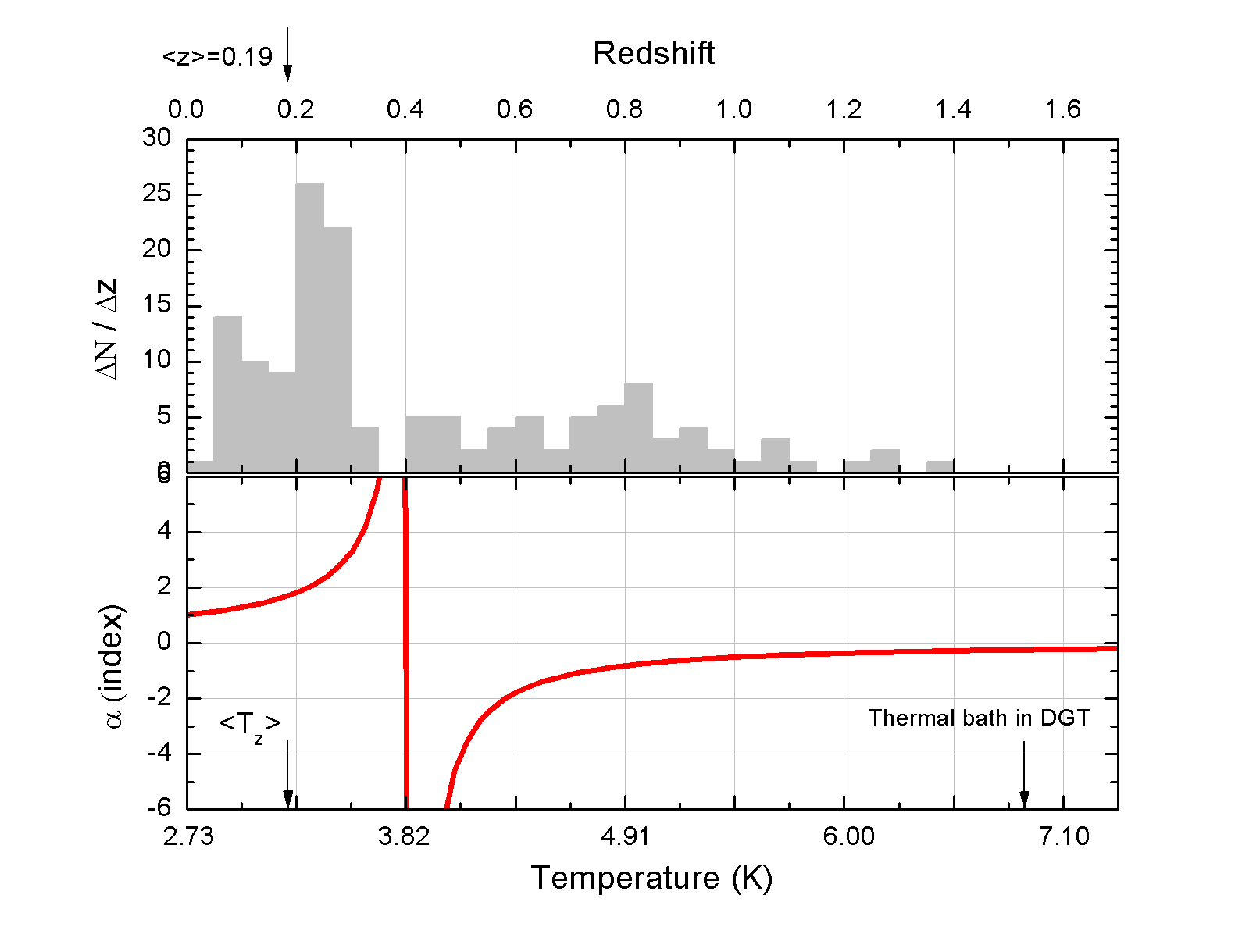}
\vspace*{-0.0cm}
\caption{Top panel: distribution of the redshift from 146 galaxy clusters. The data were compiled by \cite{reic11}. Bottom panel: dependence of the index $\alpha$, according to
 Eq.~\ref{alpha_a} of DGT. The two vertical arrows at the left indicate the average redshift of the galaxy clusters and it associated average temperature, while the vertical arrow at the right indicates the temperature value that DGT needs to describe the M-$\sigma$ and M-Tx relations of the galaxy clusters up to redshift $z\sim 0.4$.
}
\label{alpha_redshift}
\end{figure}

Fig.~\ref{alpha_redshift} (bottom panel) shows the dependence of the index $\alpha$  with the temperature (lower horizontal axis) and redshift (upper horizontal axis) according Eq.~\ref{alpha_a}. For comparison, we have included the distribution of the redshift,  from 146 galaxy clusters in the top panel. The data were compiled by \cite{reic11}, and they claim that this high-quality data, constrain the redshift evolution of X-ray scaling relations of galaxy clusters out to $z\sim 1.5$.
 
The data clearly shows two groups of galaxy clusters, those with redshift less than 0.4 and they are the majority in this data, and there is a second group of clusters, those with redshift above 0.4. The separation in redshift between these two groups at around $z\sim 0.4$ and is marked for an absence de clusters in the data and coincides with the discontinuity observed in the dependence  of the $\alpha$ index with the redshift (temperature).

Considering only the first group of clusters up to $z\sim 0.4$, they have an average redshift value of $<z>=0.19\pm 0.08$(top vertical arrow in Fig.~\ref{alpha_redshift}) and is expected that the average environment temperature, i.e., the thermal bath where they
are  immersed, must have in average a temperature of $T_z=3.25$ K (left bottom vertical arrow in Fig.~\ref{alpha_redshift}), and correspond to a index $\alpha \sim 1.65$. However, as will be shown in section~\ref{xray_emission}, the analysis of galaxy clusters in DGT up to $z\sim 4$) requires an index $\alpha=-0.24$ and correspond to a temperature of 6.92 K (right bottom vertical arrow in Fig.~\ref{alpha_redshift}. 
This discrepancy, in the environment temperature of galaxy clusters, will be discussed in section~\ref{xray_emission}.

 

\subsection{Generalized Faber-Jackson relation for galaxy clusters}
\label{faber}

Galaxy clusters have no clearly defined natural outer boundary. One way of determining the size, that is, the cluster radius is to establish this radius in such a way that it describes the same corresponding boundary for clusters of all sizes in the framework of the self-similar cluster structures \citep{kais86}. 
For instance, the $r_{\Delta}$  is the cluster radius within which the enclosed average mass density is $\Delta$
times higher than the universe critical density $\rho_c$.
\begin{equation}
M_{\Delta}=\Delta \times \frac{4}{3}\pi\;r_{\Delta}^3\;\rho_c.
\label{m_200}
\end{equation}

In DGT the application of the Eq.-\ref{eq:mainDGT} to the galaxy clusters requires careful analysis because galaxy clusters are three-dimensional stellar systems supported, in some cases, more by random motions than organized rotation.  
  Under the assumption of spherical symmetry and following the Eq.~\ref{eq:mainDGT}, the asymptotic $(r \rightarrow r_{\Delta})$, allow us calculate the mass of cluster as
\begin{equation}
M_{\Delta}(r\rightarrow r_{\Delta})= \frac{r_{\Delta}^2}{G} a (\frac{a}{a_0})^{\alpha}.
\label{eq:massa_c}
\end{equation}

For a group of objects, such as galaxies forming an open cluster, only the line-of-sight velocities are obtained, measuring the Doppler width of spectral lines of a collection of objects. In general, the line-of-sight velocity of a cluster decrease as the distance to the center increase, forming a caustic structure.  Thus, the relevant information in a cluster is the line-of-sight velocity dispersion 
$\sigma=<v^2>^{1/2}$, of galaxies in the cluster. Under certain conditions, the measurement of the velocities dispersion permits obtaining the cluster's mass from Virial theorem.

In the 7's \cite{fabe76} found a correlation, an empirical power-law relation between the luminosity of an elliptical galaxy and the velocity dispersion of its stars, expressed as $L\propto \sigma^{\gamma}$, where the index $\gamma$ is a number close to 4.
The Faber-Jackson relation is very similar to the Tully-Fisher relation, where the circular velocity observed in spiral galaxies is replaced by the dispersion velocity of the elliptical galaxies. In this sense, the acceleration can be written as $a=\sigma^2/r$
to obtain through Eq.~\ref{eq:massa_c} the generalized version of the Faber-Jackson relation predicted by DGT as
\begin{equation}
M_{\Delta}(r\rightarrow r_{\Delta})= \frac{r_{\Delta}^{1-\alpha}}{G a_0^{\alpha}} \sigma^{2\alpha+2}.
\label{faber_dgt}
\end{equation}
We can see that for $\alpha=1$, the generalized Faber-Jackson relation is a scaling-law
\begin{equation}
M_{\Delta}(R\rightarrow r_{\Delta})= \frac{1}{G a_0} \sigma^4.
\end{equation}

In \cite{sand94} there is an application of the Faber-Jackson relation (scaling-law) describing galaxy cluster within the MOND paradigm.
In this paper, we will try to a more embracing description of galaxy clusters from the Eq.~\ref{faber_dgt},  the so-called DGT version of Faber-Jackson relation, or maybe with more propriety called as, the generalized Faber-Jackson relation.

\subsection{The M-$\sigma$ relation}
\label{m_sigma}

The cluster mass estimation through the virial theorem requires that the galaxies member of the cluster and their surrounding gas must be in hydrostatic equilibrium, or at least near it.  To avoid this limitation is possible to determine a radius of virialization within which the galaxies are in a relaxing regime.
In contrast with the virial method, the gravitational lensing and the caustic technique as cluster mass estimators do not rely on the equilibrium assumption. However, at a large radius, the gravitational lensing is contaminated by line-of-sight structure unrelated to the cluster \citep{hoek11}, already the caustic technique assumes that only that galaxies trace the velocity field \citep{falt05}. In some clusters, at moderate redshift, the caustic masses agree with weak lensing masses \citep{diaf05}, and in general, the caustic mass profiles near the virial radius are consistent with the virial mass profiles.

\begin{figure}
\vspace*{-0.0cm}
\hspace*{0.0cm}
\centering
\includegraphics[width=18.0cm]{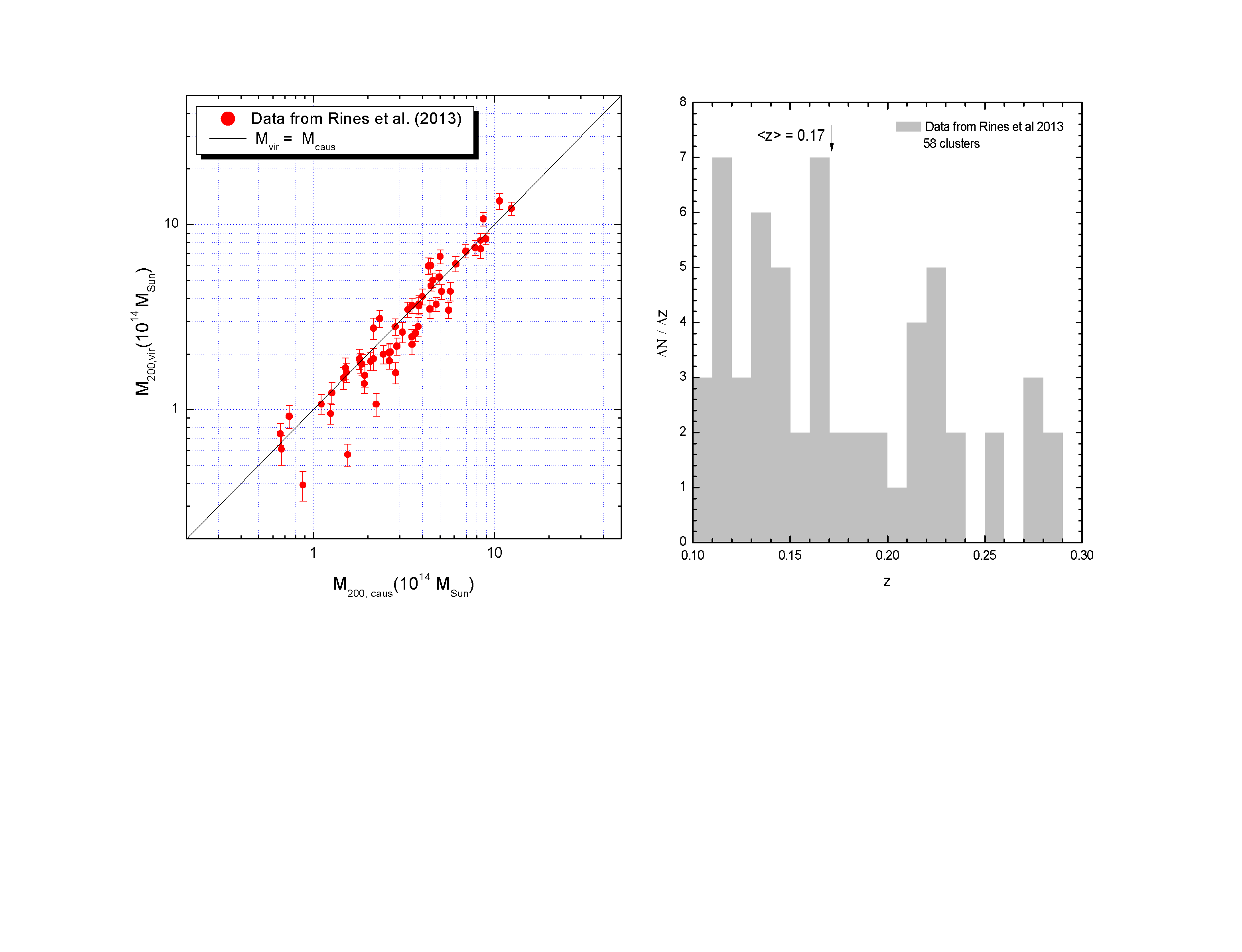}
\vspace*{-4.0cm}
\caption{Left panel: Caustic masses at $r_{200}$ (determined from the caustic
mass profile) compared to virial masses at the same radius, 
according to the data from \cite{rine13}.
Solid line has slope unity. Right panel: Redshift distribution of 58 galaxy clusters, according to the data from \cite{rine13}.
}
\vspace*{-0.0cm}
\label{Mvir_M200}
\end{figure}

Data from Cluster Infall Regions in the Sloan Digital Sky Survey (CIRS) project and the Hectoscpec Cluster Survey (HECS) project \citep{rine13} show that the caustic mass is slightly larger than virial mass at the same radius. However, the linear correlation between them is better to masses of clusters above $2\times 10^{14}M_{Sun}$. While the dispersion between them increases as the cluster mass decrease, as shown in Fig.~\ref{Mvir_M200} (left panel).

\begin{figure}
\vspace*{+0.0cm}
\hspace*{0.0cm}
\centering
\includegraphics[width=15.0cm]{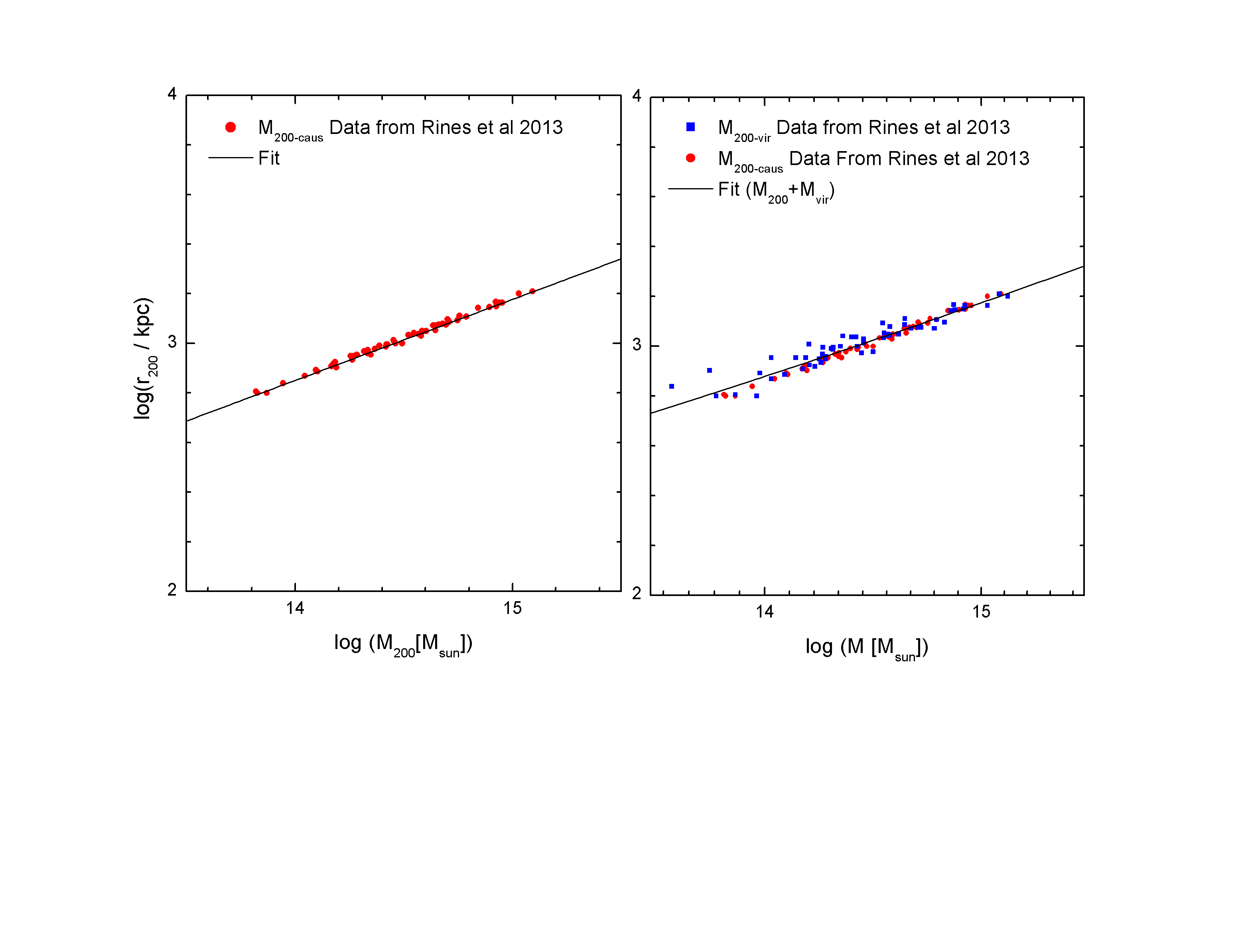}
\vspace*{-3.0cm}
\caption{Left panel: Red circles, $r_{200}-M_{200-caustic}$ correlation. Data from \citep{rine13}. The solid line is a least squares fit.
Right panel: Blue squares $r_{200}-M_{200-virial}$ correlation.
Red circles  $r_{200}-M_{200-caustic}$ correlation. Data from \citep{rine13}. The solid line is a least squares fit.
}
\vspace*{-0.0cm}
\label{r_m}
\end{figure}

Also,  the HeCS data represent a cluster population with a wide range of redshift, from $z=0$ to $z=0.3$. Fig.~\ref{Mvir_M200} (right panel) show the redshift distribution. According to data compiled by \cite{reic11}, galaxy clusters up to redshift($\sim$ 0.4) is consistent with none evolution
with the redshift.

An important features of Eq.~\ref{faber_dgt} is the dependence of cluster mass with the radius, the exception is for $\alpha=1$ (deep-MOND regime). The data is consistent with a power-law dependence between the cluster radius and the mass and expresed as
\begin{equation}
r_{\Delta}=c M_{\Delta}^b;
\label{r_m1}
\end{equation}
Fig.~\ref{r_m} shows this correlation for the data  of HECS project \citep{rine13}. The left panel shows the correlation between $r_{200}$ and the $M_{200}$ (caustic mass) expressed as $\log r_{\Delta}= \log c+b\log M_{\Delta}$ with $\log c=-1.727\pm 0.049$ and $b=0.327\pm 0.003$ when the radius is measures in kpc and the mass in solar masses ($R-S_{qua}=0.994$). While, in the right panel is also included the mass  of the clusters obtained by the virial method, in this case,
$\log c=-1.261\pm 0.107 $ and $b=0.296\pm 0.007$ ($R-S_{qua}=0.933$). We can see that the addition of the virial masses on the analysis, increase the dispersion in the r-M relation.

Substituting Eq.~\ref{r_m1} in Eq.~\ref{faber_dgt} the expression for the cluster mass, can be written as
\begin{equation}
\log \left(  M_{\Delta}\right)=A+B\log \sigma,
\end{equation}
where the normalization is given as
\begin{equation}
A=\frac{1}{1-b(1-\alpha)}\left[\log \left( \frac{c^{1-\alpha}}{Ga_0^{\alpha}}\right)\right],
\end{equation}
and the slope as
\begin{equation}
B=\frac{2(\alpha+1)}{1-b(1-\alpha)},
\end{equation}
These equations reproduce the scaling-law predicted by the deep-MOND regime ($\alpha=1$, in DGT), with $A=-\log(Ga_0)$ and slope $B=4$. However, according to \cite{milg18}, this normalization parameter would be valid to an individual member of a  galaxy group. But  for the case of a group made
of N $\gg$1 galaxies with individual masses $m_i$ and $M=\Sigma m_i$, the normalization becomes 
$A\sim \log(81/4)-log(Ga_0)$.

The $\Lambda$CDM prediction to $M-\sigma$ relation comes putting  Eq.~\ref{m_200} as a function of velocity $V_{\Delta}$ as
\begin{equation}
M_{\Delta}=(\frac{\Delta}{2})^{-1/2} (GH_0)^{-1} V_{\Delta}^3.
\label{m_200b}
\end{equation}
It is expected in $\Lambda$CDM that the circular velocity of a galaxy is in association with the peak velocity of an
NFW halos \citep{nava97}, as $V_c=f_v V_{\Delta}$ with $f_v\sim 1$ \citep{mcga10} (for $\Delta=500$). Similarly, for galaxy cluster, the relation can be extended as
\begin{equation}
\sigma \sim f_v V_{\Delta},
\label{fv}
\end{equation}
as will be shown below, for $\sigma=f_v V_{200}$, the cluster data is consistent with $f_v \sim 0.75$.
 Combining the Eq.~\ref{fv} and Eq.~\ref{m_200b} we have
the $\Lambda$CDM prediction to the $M_{200}-\sigma$ relation, parametrized as
\begin{equation}
M_{200}= (204789\; km^{-3} s^3 M_{sun}^{-1})(\sigma/0.75)^3.
\end{equation}

\begin{figure}
\vspace*{+0.0cm}
\hspace*{0.0cm}
\centering
\includegraphics[width=14.0cm]{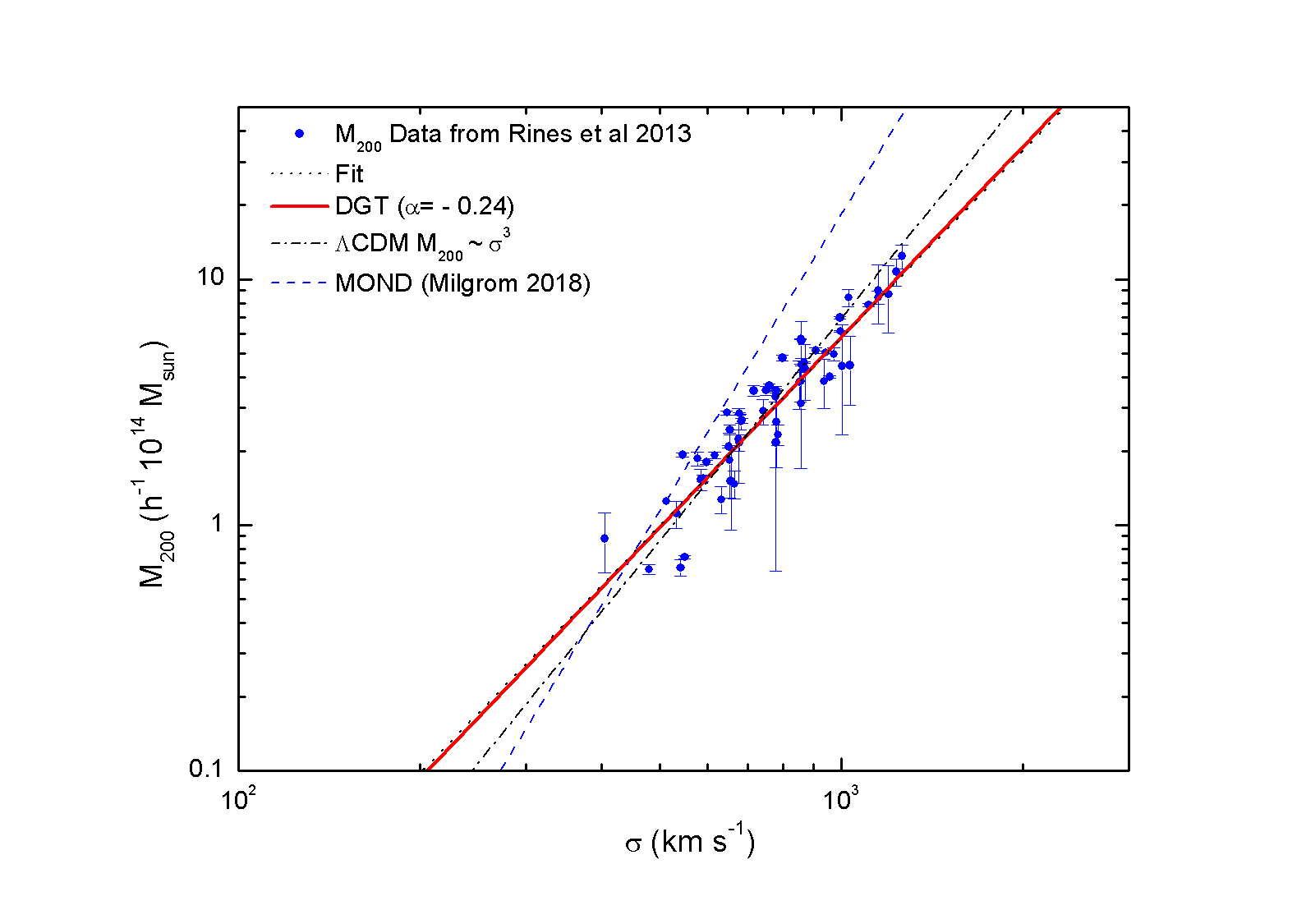}
\vspace*{-0.0cm}
\caption{Comparison of the $M_{200-caustic}-\sigma$ relation data from Hectoscpec Cluster Survey (HECS) project \citep{rine13} and several theoretical predictions (lines), including a fit on data (dots line)(h$=$0.7).
}
\vspace*{-0.0cm}
\label{m_sigma1}
\end{figure}

\begin{figure}
\vspace*{+0.0cm}
\hspace*{0.0cm}
\centering
\includegraphics[width=14.0cm]{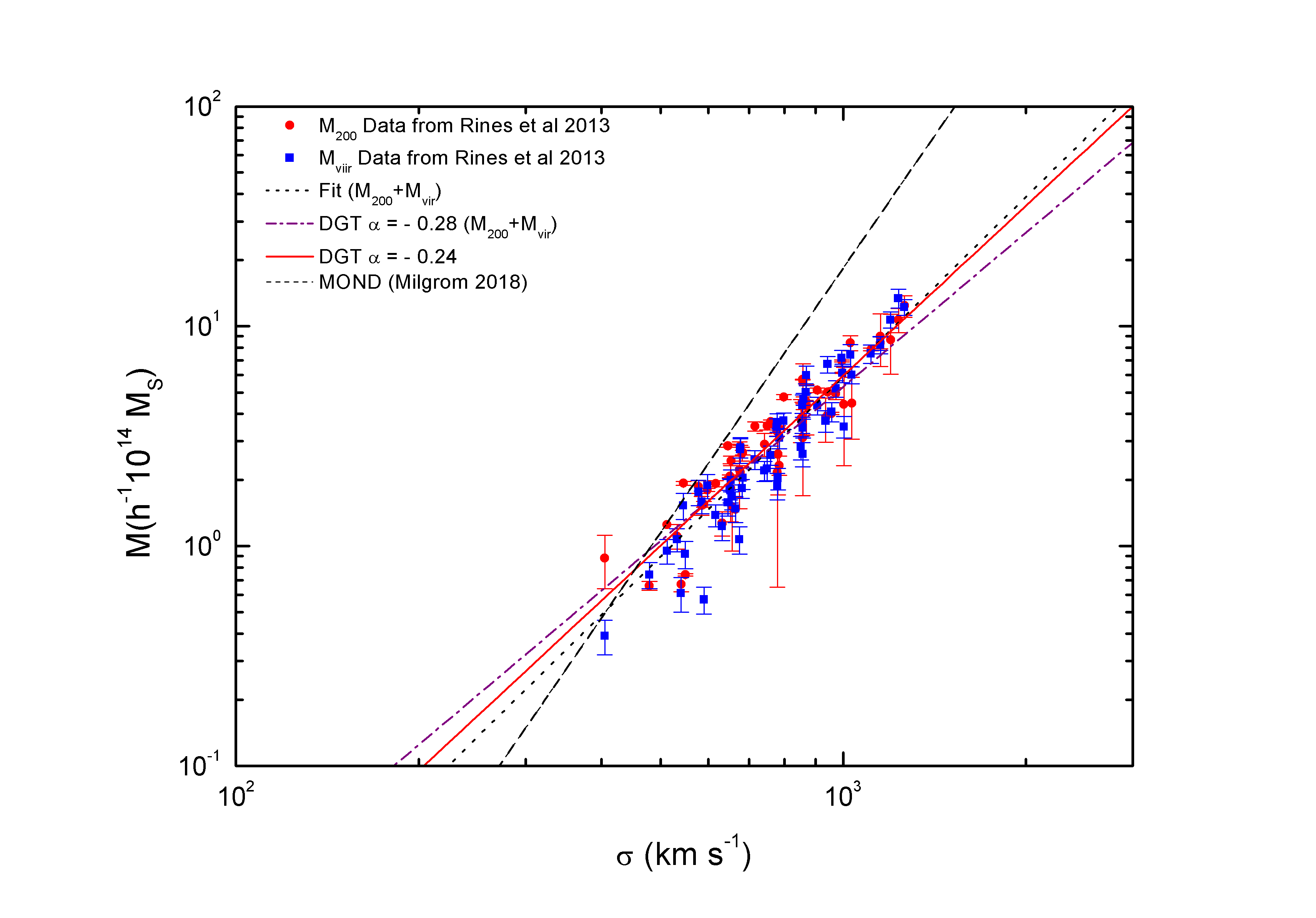}
\vspace*{-0.0cm}
\caption{Comparison of the $M_{200-caustic}-\sigma$ relation data (red circles) and 
the $M_{200-virial}-\sigma$ relation data (blue squares),
from Hectoscpec Cluster Survey (HECS) project \citep{rine13} and several theoretical predictions (lines), including a fit on data (dots line) (h$=$0.7).
}
\vspace*{-0.0cm}
\label{m_sigma2}
\end{figure}

Fig.~\ref{m_sigma1} shows the $M_{200}-\sigma$ relation, is a comparison among the data from of HECS project \citep{rine13} with DGT predictions ($\alpha =-0.24$, solid red line), including a fit of the data (dotted line), $\Lambda$CDM ($f_v=0.75$, dash dot line) and the MOND prediction \citep{milg18} (dashed line). The upper part of Table ~\ref{m_sigma_table} shows the values for the normalizations (A) and slopes (B). We can see, that the DGT prediction for $\alpha=-0.24$ is practically overlapping to the fit line.
Fig.~\ref{m_sigma2} is similar to the previous one, but including in the analysis the mass of clusters obtained via the virial method.
The lower part of Table ~\ref{m_sigma_table} indicates the values to the normalizations and the slopes. 

\begin{table}[h!]
\begin{center}
 \caption{Table with aligned units.}
 \begin{tabular}{ l  l  l  l l}
    \hline
    \hline
    Relation           &    A                &   B             &  $\chi^2$& dof=57  \\
    \hline
    $M_{200}-\sigma$   & $7.17\pm 0.37$     &   $2.53\pm 0.13$& 0.045& Fit  \\ 
                       &  6.90              &    2.57         & 0.059& DGT $\alpha=-0.24$ \\
                       &  5.69              &    3.00         & 0.067&$\Lambda$CDM  \\ 
                       &  3.11              &    4.0          &  0.52&MOND (Milgrom, 2018)\\ 
   \hline
  $M_{200,vir}-\sigma$ &  6.90              &   2.57          & 0.099 &DGT $\alpha=-0.28$\\
\hline  
     \hline                  
    \end{tabular}
    \label{m_sigma_table}
\end{center}
\end{table}

From these figures, we can see that the data constrain the MOND prediction. We believe that the fact of the cluster mass to be independent of the cluster size, as predicted by MOND impose a severe limitation for the inclination in the $M-\sigma $ relation, a slope with index always equal to 4, is constrained by several cluster surveys.

\subsection{X-ray Emission from Clusters of Galaxies}
\label{xray_emission}

According to the self-similar model \citep{kais86,bowe97}, the properties of clusters
reflect the properties of the Universe at their redshift of observation. Indeed, galaxy clusters are open systems, and the determination of their sizes is possible admitting, e.g., a radius within which the mean density is $\Delta$ times the
critical density ($\rho_c$) of the Universe, at the cluster's redshift. And defines a self-similar structure of the clusters. As $\rho_c$ is a function of the redshift, distant clusters are identical to local clusters if we
include a factor for increasing density of the Universe with
redshift. Thus, galaxy clusters surveys can give valuable information on the evolution of the Universe.

The self-similar model implicates that clusters form via a single gravitational collapse at a redshift of observation and the only source of energy input into the intracluster medium (ICM) is gravitational. 
The model predicts that the slopes in the Lx-T, M-T and M-Tx relations are independent of the redshift, and only the normalizations have an evolution with the redshift, for instance, the normalization in the M-T relation can be written as \citep{reic11}

\begin{equation}
\frac{M_{obs}}{M_{z=0}(T)}=E(z)^{\alpha_e},
\end{equation}
where the $E(z)$ is the evolution function and the index as $\alpha_e=-1.0$, is predicted by the self-similar model. E(z) is an increasing function of z that depends on
cosmological parameters (e.g. $\Omega_M$, $\Lambda$).

This evolution is hard to see in the data at redshift range ($z< 6$), it is claimed that in this redshift range, the data has a strong influence by selection bias, constraining the evolution observations \citep{reic11}, in short; there is not an apparent evolution in the data. In contrast to the more distant systems ($z>6$), there is an evolution in the data, but smaller than the predicted by the self-similar model. 
This behavior also suggests that some non-gravitational processes can be responsible for the absence of the evolution, at least for galaxy clusters with redshift up to $z \sim 4$.

We show a scenario to describe X-ray galaxy clusters, without an apparent evolution within DGT, assuming that the X-ray emission, results
from thermal bremsstrahlung from a hot diffuse intracluster gas \citep{felt66}.

In DGT, all systems are within a thermal bath, and the dynamic of a system depends on the temperature of the thermal bath. In general, if a system is a ``isolated'' galaxy at redshift z, its thermal bath has a temperature of $T=T_0(1+z)$, where $T_0=2.73$ K.
For example, if the system is an isolated nearby spiral galaxy ($z\sim 0$), it is within a thermal bath at 2.73 K, and this corresponds to ($\alpha =1$) in the DGT equations.
Already, if the system is a dwarf galaxy orbiting a nearby spiral galaxy, its thermal bath has a temperature slightly higher than 2.73 K, because it is subject to additional radiation of its host galaxy \citep{navi18b}. In DGT, dwarf galaxies are described by an index $\alpha > 1$. 
 
So if the system is a ``isolated'' galaxy cluster at a redshift, let's say $z\sim 0.17$ (this is the average redshift in the HECS cluster survey \citep{rine13}), 
its thermal bath at this redshift must have a temperature of 3.19 K. However, the analysis of the M-$\sigma$ relation in DGT require an index $\alpha=-0.24$, as shown in the previous section, and according to the $\alpha$-T diagram of DGT (Fig.~\ref{alpha_redshift}), $\alpha=-0.24$ correspond thermal bath with a temperature of 6.92 K.
This discrepancy can be understood assuming that the X-ray emission mechanism is the thermal bremsstrahlung from intracluster gas, as follow. 


An exponential behavior dominates the X-ray spectra in galaxy clusters; this means a thermal X-ray emission, with gas temperatures about $10^8$ K. However, in some clusters, there is a contamination
of a power-law spectrum, this means, a no thermal origin.
The origin of this component is due to some individual galaxies within the cluster, such as active galactic nuclei (AGN),  or even binary stellar X-ray sources. Also, there are some emission lines, whose origin can be linked to the contamination of the gas by heavy elements.

\begin{figure}
\vspace*{-0.0cm}
\hspace*{0.0cm}
\centering
\includegraphics[width=12.0cm]{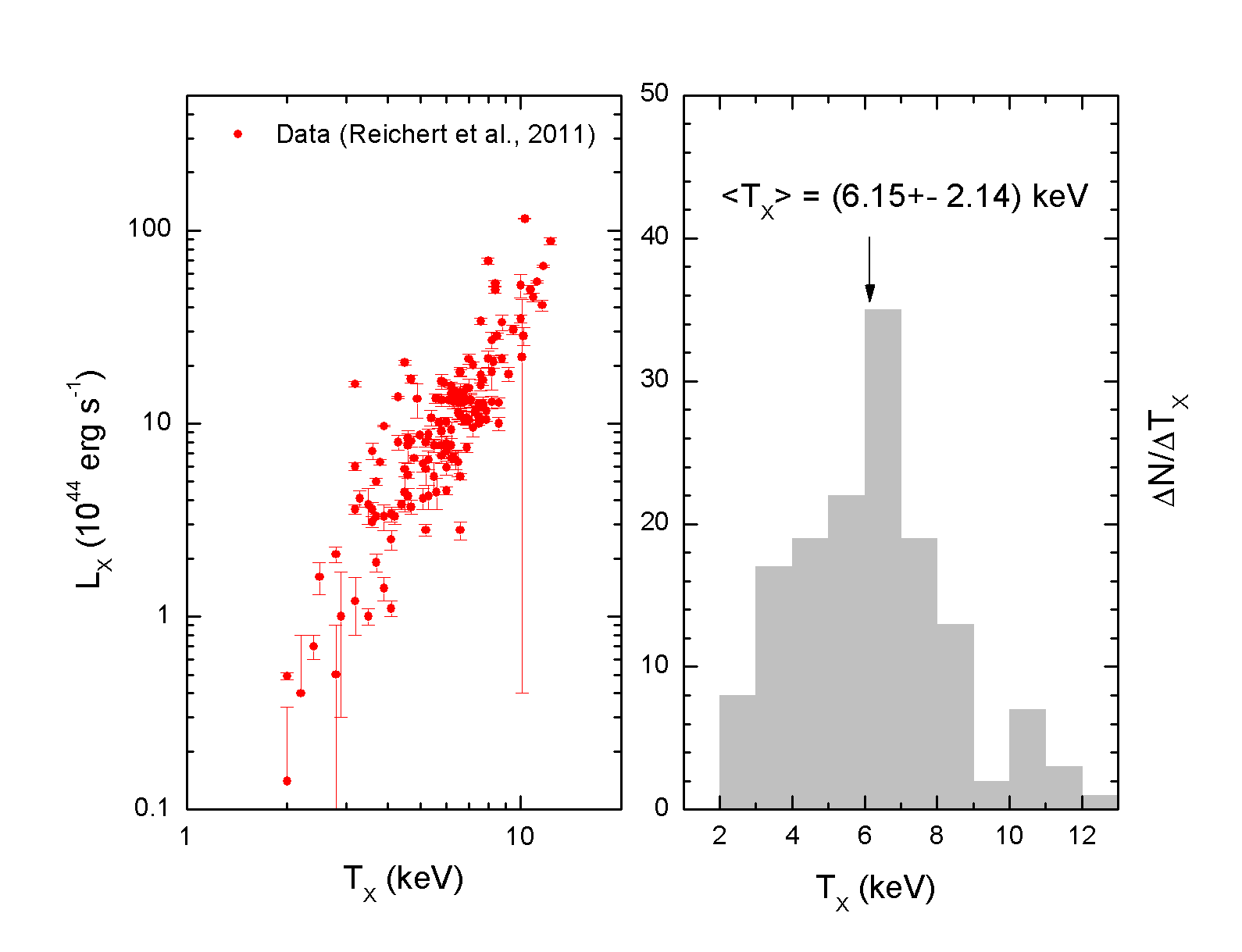}
\vspace*{-0.0cm}
\caption{Left panel the Lx-Tx relation and right panel the  ICM temperature distribution, obtained from 148 galaxy clusters according to data compiling by \cite{reic11}.
}
\label{average_T}
\end{figure}

The gas temperature that fit the exponential spectrum in an X-ray cluster is about $2\times 10^7$ to $10^8$ K. \citep{felt66,sara88}
These temperature values are in agreement with the average temperature $T_X=6.15 \pm 2.14$ K, obtained from 148 galaxy clusters, 
as shown in Fig.~\ref{average_T} (right panel), together with the $L_X-T_X$ relation (left panel) and whose data was compiled by \cite{reic11}. The X-ray clusters, have luminosities from $10^{43}$ to $10^{45}$ ergs sec$^{-1}$ and constituting the most common and bright extragalactic X-ray sources. 

The cluster gas temperature is close to the thermal dispersion velocity of the particles (mostly protons) of the gas
\begin{equation}
 \sigma^2 \sim \frac{T_X}{\mu m_p},
 \label{thermal_2}
\end{equation} 
where $m_p=9.35\times 10^5$ keV is the proton mass, and $\mu=0.62$ is the is the mean atomic weight (for a fully ionized gas
with solar abundances).

The gas temperatures around $Tx\sim 10^8$ K or 8.6 keV can provide high thermal velocities to the gas particles, sometimes above
of the escapement velocity of the gravitational attraction of the cluster and a fraction of the gas would escape as a wind. A favorable condition to this mechanism it's the clusters are open systems. Notice, that the atomic density of the hot gas that fills an X-ray cluster is about
$n\sim 10^{-3}$ cm$^{-3}$, even so, the total mass of the gas can exceed the mass of stars of the galaxies within the cluster.

This hot wind can warm the local environment, the thermal bath, where the cluster is immersed. This mechanism can put all the galaxy clusters immersed in thermal baths, with almost the same temperature $T\sim 6.92$ K (which correspond to $\alpha=-0.24$), independent of the cluster redshift,
and explain the weak (or nearly absent) evolution with the redshift, in the galaxy clusters relations at least up to redshift $z\leq 0.4$.


\subsection{The $M-T_X$ relation}
\label{m_t}

In most cases, the hot gas of galaxy clusters are described by the hydrostatic equation [Sarazin] and that under a spherical symmetry, can be written as
\begin{equation}
\frac{1}{\rho_g}\frac{dP}{dr}=-\frac{d\phi (r)}{dr}=-\frac{GM(r)}{r^2},
\end{equation}
where P is the gas pressure, $\rho_g$ is the gas density, $\phi (r)$ is the gravitational potential and $M(r)$ is the total cluster mass within r. The next step is to determine an expression to describe the density distribution of hot gas in clusters. The so-called "$\beta$ model" (or pure gas sphere) allows obtaining a reasonable description of this density distribution.
Also, for large values of $r$ is possible to obtain an asymptotically simple expression.

However, in this work and to avoid free parameters and asymptotic approximations, we will use to obtain the $M-T_X$ correlation of galaxy clusters a different frame from the DGT version to the Faber-Jackson relation, express by Eq.~\ref{faber_dgt},
\begin{equation}
M_{\Delta}(r\rightarrow r_{\Delta})= \frac{r_{\Delta}^{1-\alpha}}{G a_0^{\alpha}} \sigma^{2\alpha+2},
\label{faber_dgt2}
\end{equation}
and taking into account the Eq.~\ref{thermal_2}, that gives the relationship between the hot gas temperature of a cluster, $T_X$, and the velocity dispersion, $\sigma$, \citep{sara88}.
\begin{equation}
T_X=6.03\;keV\;\left[\frac{\sigma}{10^3\;km\;s^{-1}}\right]^2.
\label{thermal}
\end{equation}

In the generalized Faber-Jackson relation the M-$\sigma$ relation is a power-law such $M\propto \sigma^{2(\alpha+1)}$. Thus Eq.~\ref{thermal} can be rewritten as

\begin{equation}
\left(\frac{\sigma}{km\;s^{-1}}\right)^{2(\alpha+1)}=10^{6(\alpha+1)}\left(\frac{T_X}{6.03\;keV} \right)^{\alpha+1}.
\label{sigma_T}
\end{equation}

Considering that the size (radius) of cluster is like $r_{\Delta}=c M_{\Delta}^b$, for instance, the $r_{500}$  is the cluster radius within which the enclosed average mass density is 500 
times higher than the universe critical density $\rho_c$. Data the ROSAT All-Sky Survey \citep{ande15} can be expressed as power-law
\begin{equation}
\log(r_{500})= \log(c)+b \log M_{500},
\label{r_500}
\end{equation}
with $\log(c)=-1.725 \pm 0.09)$ and $b=0.323 \pm 0.001$ when $r_{500}$ is expressed in kpc and $M_{500}$ in solar masses. 

Combining the Eq.~\ref{r_500}, Eq.~\ref{sigma_T} and Eq.~\ref{faber_dgt2} 
the expression to mass of cluster can be written as 
\begin{equation}
\log \left( M_{500}\right) =A+B \log\left(\frac{T_X}{keV}\right),
\end{equation}
where the normalization factor is giving by
\begin{equation}
A=\frac{1}{1-b(1-\alpha)}\left[\log\left(\frac{c^{1-\alpha}}{Ga_0^{\alpha}}\right)+(6-\log 6.03)(\alpha+1)\right],
\end{equation}
and the slop as
\begin{equation}
B=\frac{(\alpha+1)}{1-b(1-\alpha)}.
\end{equation}
 These equations include the special case predicted by the deep-MOND regime ($\alpha=1$, in DGT). In this case the normalization factor is reduce to 
\begin{equation}
A=-\log(Ga_0)+10.44=12.71.
\end{equation}
and the slop as $B=2$. However, this normalization factor differ from an isothermal gas sphere in MOND analysis \citep{sand03}, that predict a higher normalization factor as
$A=13.46$

\begin{figure}
\vspace*{+0.0cm}
\hspace*{0.0cm}
\centering
\includegraphics[width=14.0cm]{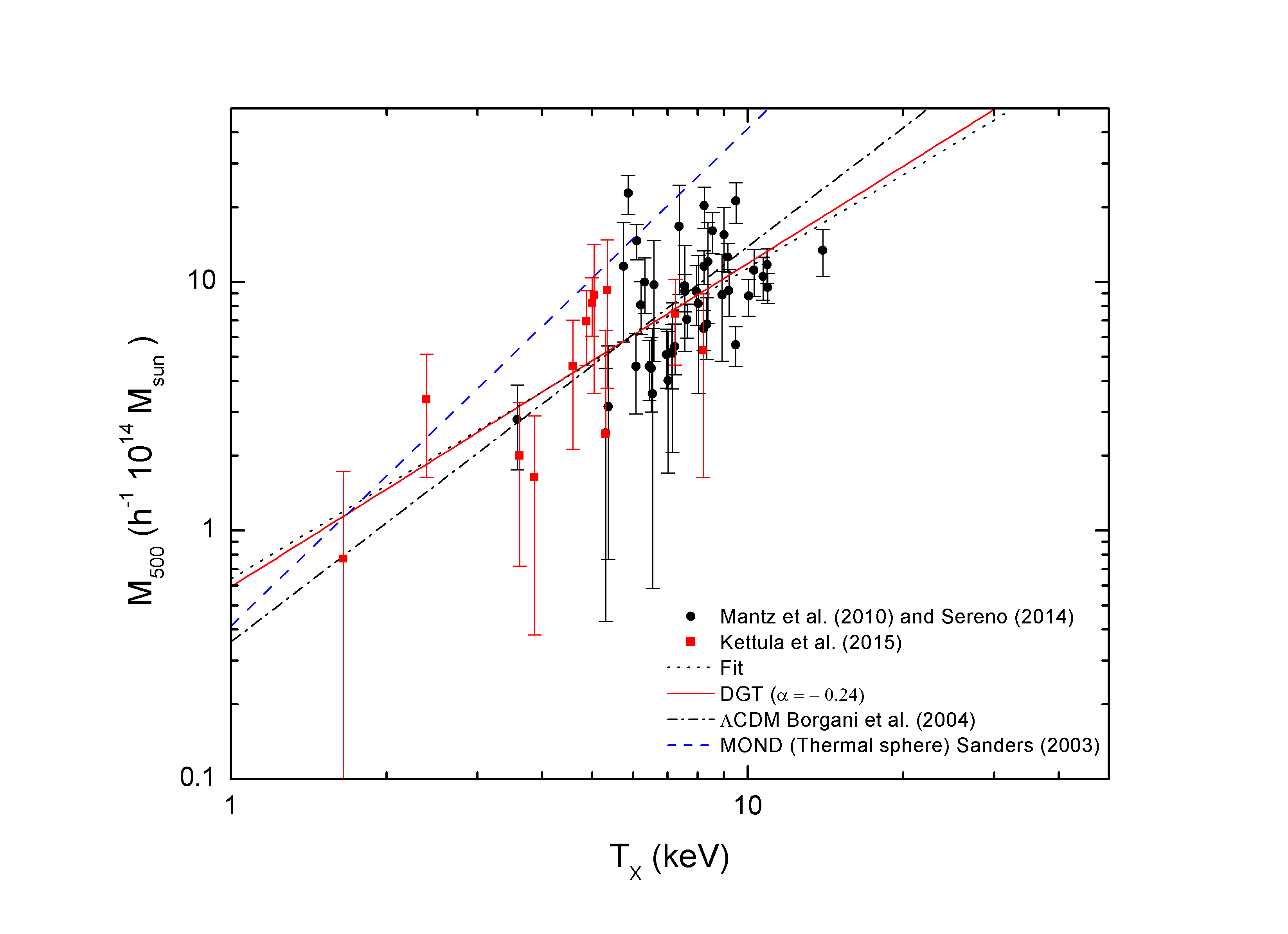}
\vspace*{-0.0cm}
\caption{$M_{500}$-Tx relation, data  from 
Canada–France–Hawaii Telescope Lensing Survey and XMM-CFHTLS survey up to $z\sim 0.47$. Red squares from \cite{kett15}  and black circles from \cite{mant10} and \cite{sere14}. The lines represent several theoretical predictions, including a fit on data (dots line) 
(h$=$0.7).
}
\vspace*{-0.0cm}
\label{m_Tx}
\end{figure}
 
 Fig.~\ref{m_Tx} shows the $M_{500}-T_X$ relation, is a comparison among the data from Canada–France–Hawaii Telescope Lensing Survey and XMM-CFHTLS surveys up to $z\sim 0.47$. \citep{kett15} with DGT predictions (solid red line), including a fit of the data (dotted line) and the MOND prediction (thermal sphere) \citep{sand03} (dashed line). Table ~\ref{m_Tx_table} shows the values for the normalizations (A) and slopes (B). We can see, that the DGT prediction for $\alpha=-0.24$ is practically overlapping to the fit line.
 
\begin{table}[h!]
\begin{center}
 \caption{Table with aligned units.}
 \begin{tabular}{ l  l  l  l  l}
    \hline
    \hline
    Relation           &    A                &   B             &  $\chi^2$  & \\
    \hline
    $M_{500}-T_X$   & $13.80\pm 0.15$     &   $1.25\pm 0.18$& 0.16 & Fit  \\ 
                       &  13.62             &    1.30        & 0.27& DGT $\alpha=-0.24$ \\
                       &  13.75             &    1.54         & 0.36&$\Lambda$CDM (Borgani et al. 2004)\\
                       &  13.46             &    2.00          & 2.39& MOND (Thermal sphere)Sanders, 2003 \\ 
  \hline  
     \hline                  
    \end{tabular}
    \label{m_Tx_table}
\end{center}
\end{table}

\section{A toy model to the caustic pattern in the galaxy cluster}
\label{toy}

In a group of objects, such as galaxies forming an open cluster, the line-of-sight velocity is obtained by measuring
the Doppler width of spectral lines of a collection of objects. In general, the line-of-sight velocity (in the clustercentric rest-frame) decrease as the
distance to the center increases, forming a Caustic structure. 

The Caustic technique \citep{diaf05,rine13}
 determines the line-of-sight velocity, as a function of the cluster projected radius, forming the so-called phase space, line-of-sight velocity vs. projected radius. The edge of this phase space traces the galaxy escape velocity from the group,  to obtain the Newtonian gravitational potential through the relation 
\begin{equation}
v_{esc}^2=-2\phi(r), 
\label{escape}
\end{equation}
and from it, can estimate the cluster mass \cite{diaf05}.
The gravitational potential $\phi(r)$ is related with the radial acceleration by the expression
\begin{equation}
\phi(r)=\int a(r)dr.
\label{potential}
\end{equation}
From Eq.~\ref{eq:mainDGT} the radial acceleration in DGT is 
\begin{equation}
a(r)=\frac{(GMa_0^{\alpha})^{1/(\alpha+1)}}{r^{2/(\alpha+1)}},
\label{acceleration}
\end{equation}
as expected, for $\alpha=0$, the above expression coincides with the Newtonian radial gravitational acceleration  $a(r)=GM/r^2$.
Incorporating Eq.~\ref{acceleration} into Eq.~\ref{potential} and integrating, we have a expression for the radial gravitational  potential predicted by DGT as
\begin{equation}
\phi(r)=(GMa_0^{\alpha})^{1/(\alpha+1)}  \left(\frac{\alpha+1}{\alpha-1}\right)r^{(\alpha-1)/(\alpha+1)},
\label{potential}
\end{equation}
again, as expected, for $\alpha=0$ the above potential, coincides with the Newtonian gravitational potential as $\phi(r)=-GM/r$

 Our toy model take on that $v_{esc}(r)=\pm \sqrt{-2\phi(r)}$ are the upper and lower curves, of the envelop for the cluster velocities dispersion, generated through relation 
 $u \times \pm \sigma(r)$, where $u$ is a random number between 0 and 1, and $\sigma(r)$ is the velocity dispersion, here obtain from the generalized Faber-Jackson relation, Eq.~\ref{faber_dgt}.

\begin{figure}
\vspace*{-0.0cm}
\hspace*{0.0cm}
\centering
\includegraphics[width=12.0cm]{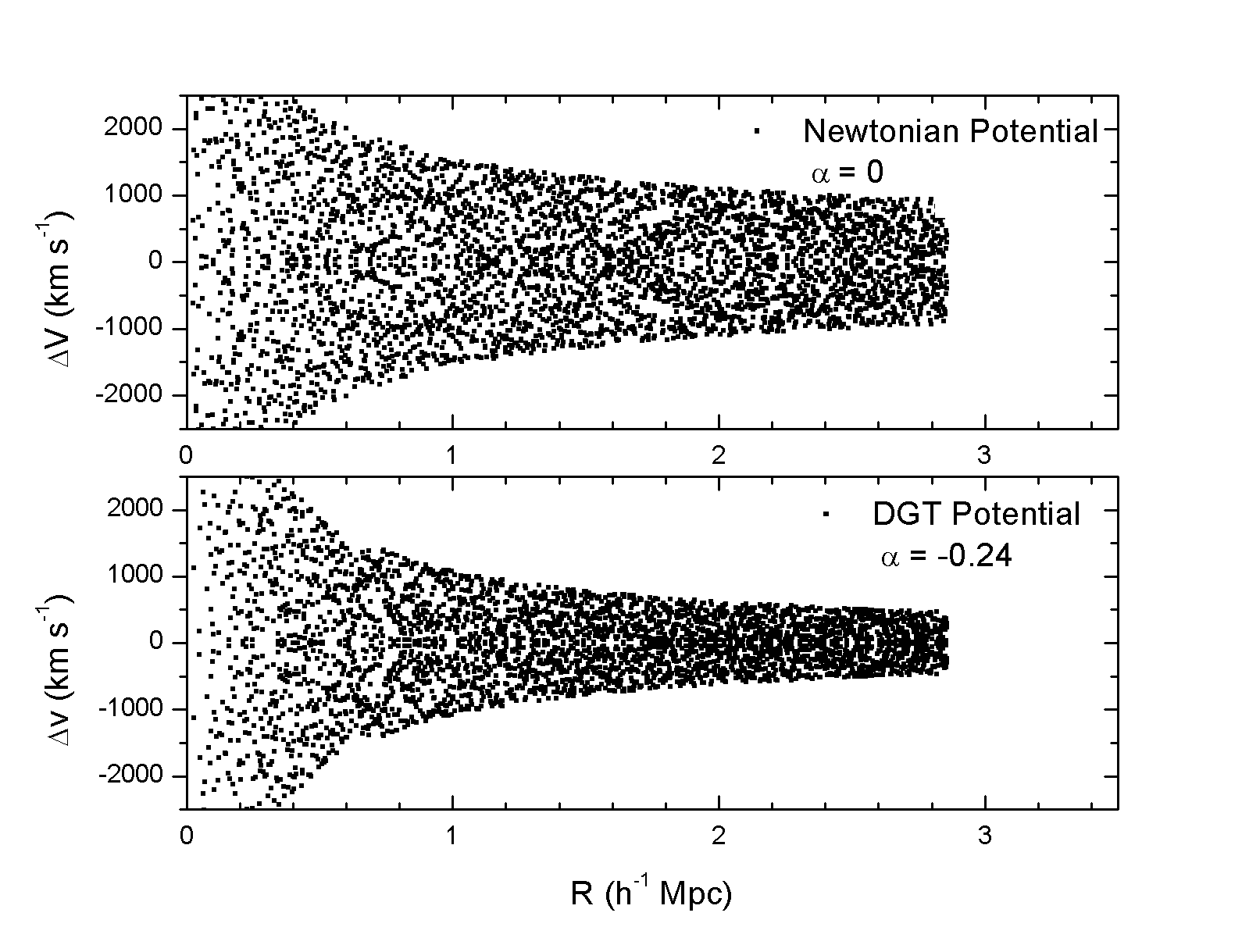}
\vspace*{-0.0cm}
\caption{Prediction for the dispersion velocities versus  projected radius, for a galaxy cluster with $M=2.0\times 10^{14} M_{sun}$. The edges follow the caustic form, under two assumptions to the galaxies escape velocity: in the top panel we assume a Newtonian gravitational potential ($\alpha=0$ in DGT) and the bottom panel we assume a DGT gravitational potential (Eq.~\ref{potential}) with $\alpha=-0.24$.
}
\label{caustic}
\end{figure}

Fig.~\ref{caustic} shows an 
example, for a cluster with mass $M=5.0\times 10^{14} M_{sun}$. The top panel takes on a Newtonian potential
($\alpha=0$ in DGT) and the bottom panel takes on a DGT gravitational potential (Eq.~\ref{potential}) with $\alpha=-0.24$.

The Caustic structure in the data show a variety of forms \cite{rine13}. However, the average behaviour are close to the DGT prediction.

 The caustic edge curves in DGT are like the declining rotation curves observed in galaxies at high redshift. DGT predicts that both, the caustic edges in galaxy clusters, and the rotation curves of galaxies at high redshift falling faster than the Newtonian prediction \citep{navi17,navi18a}. Also in both cases, a falling behavior is predicted by DGT,
only when the index $\alpha$ is negative.
In the case, of galaxy clusters, the M-$\sigma$ and M-Tx relations are well described by DGT with an index $\alpha=-0.24$ (see subsection~\ref{m_sigma} and subsection~\ref{m_t}).

\section{Conclusions}
\label{conclusions}

We have presented a novel framework, within DGT picture, that explains the almost absence of evolution with the redshift of the galaxy cluster, at least up to $z\sim 0.4$, DGT is based in an extension to low temperatures, of entropic gravitational theory \citep{verl11}. DGT allows obtaining a generalized version of the Faber-Jackson relation to describe the galaxy clusters relation, such as the M-$\sigma$ and M-Tx relations. The comparison with the data available in the literature, from large galaxy clusters surveys, shows an excellent agreement between them. 

Our main result is to show that DGT can make a description of galaxy clusters, without invoking dark matter,
always within the same framework already used to describe the dynamic of dwarf galaxies including the dwarf spheroidal galaxies orbiting the Milke-Way galaxy \citep{navi18b}, as well as, the description of the falling rotation curves, observed by VLT \citep{lang17} at high redshift \citep{navi18a}.

The weak (almost null) evolution with the redshift, in the normalization relations of galaxy clusters in the available data, constrain the simple self-similar predictions, at least up to $z\sim 4$. According to DGT, this behavior is due to
a fraction of the hot gas that fills the clusters is not bound to the groups, escaping as a hot wind and warming the clusters environment, so the clusters are within thermal baths with almost the same effective temperature ($Tx\sim 6.92$ keV or $\alpha=-0.24$) independently of the redshift of the clusters. 
DGT predictions to $\alpha=-0.24$ to the M-$\sigma$ and M-Tx relations overlap the fit on data.

Prediction of DGT for the slope ($B=1.30$) of the $M_{500}$-Tx relation is close to the slope ($B=1.25$)  of the fit on Canada–France–Hawaii Telescope Lensing Survey and XMM-CFHTLS surveys (up to $z\sim 0.47$). Both, are minors than what is predicted by self-similar models ($B=1.5$) and suggests that simple gravitational collapse is not the only process that governs the heating or cooling of the gas in clusters and their local environments. 

Also, $\Lambda$CDM prediction gives a good description of the M-Tx relation. However,  the slope ($B=1.55$) is steeper than the fit on the data ($B=1.25$). Besides,
as point out by \cite{sand03}, the mass predicted by
MOND in the M-Tx relation is a factor of 2 or 3 times larger than the observed. While to describe the inner region of galaxy clusters, MOND needs unseen matter. 

Finally, we already have shown that DGT, allow a description of galaxies, departing from the local Universe  ($z\sim 0$), 
up to redshift $\sim 4$, where the linear relation ($T=T_0(1+z)$) between the redshift and temperature of the Universe is guaranteed, as well as, the dwarf galaxies, including of local dwarf galaxies. Now, we show that DGT can describe the galaxy clusters, always with the same framework (without the dark matter paradigm). We are at the beginning, but we already have demonstrated that DGT is a promising theory that can be extended to higher redshift, maybe up to reach the redshift of the origin of the CMB radiation, and it, is our challenge.

\acknowledgments

This work is supported by the Conselho Nacional de Desenvolvimento Cient\'{i}fico e Tecnol\'{o}lgico (CNPq, Brazil, grants 312066/2016-3, 152050/2016-7, 406331/2015-4),


\end{document}